\documentclass[12pt]{article}
\usepackage[T1]{fontenc}
\usepackage{newtxtext,newtxmath}
\usepackage[utf8]{inputenc}
\usepackage[linktocpage=true,plainpages=false]{hyperref}
\usepackage{color}
\usepackage{amsmath,amssymb}
\usepackage{cite}
\usepackage{array}
\usepackage[labelsep=period]{caption}
\usepackage[title]{appendix}

\definecolor{linkcolor}{rgb}{0.6,0,0}
\definecolor{citecolor}{rgb}{0,0.6,0}
\definecolor{urlcolor}{rgb}{0,0,0.9}
\hypersetup{colorlinks, linkcolor={linkcolor},citecolor={citecolor}, urlcolor={urlcolor}}

\newcommand{\ls}{\left(}
\newcommand{\rs}{\right)}
\newcommand{\al}{\alpha}
\newcommand{\be}{\beta}

\newcommand{\dd}{\partial}
\newcommand{\de}{\delta}
\newcommand{\m}{\mu}
\newcommand{\n}{\nu}
\newcommand{\ga}{\gamma}
\newcommand{\la}{\lambda}
\newcommand{\ta}{\tau}
\newcommand{\ka}{\varkappa}

\newcommand{\disn}[2]{$$\displaylines{\refstepcounter{equation}%
		\label{#1}\hskip 1em minus 1em #2\hfilneg}$$}
\newcommand{\nom}{\hfil\hskip 1em minus 1em (\theequation)}
\begin{document}
		\title{Polyakov-like approach to the modified gravity \\ and other geometric theories}
		\author{A.~A.~Sheykin\thanks{E-mail: a.sheykin@spbu.ru}, M.~V.~Ioffe\thanks{E-mail: ioffe000@gmail.com}, S.~N.~Manida\thanks{E-mail: s.manida@spbu.ru}, S.~A.~Paston\thanks{E-mail: pastonsergey@gmail.com}\\
		{\it Saint Petersburg State University, Saint Petersburg, Russia}}
		
		\date{}
		\maketitle

	\begin{abstract}
		We study the procedure that allows to rewrite the actions of some modified gravity theories like mimetic gravity and Regge-Teitelboim embedding theory as Einstein-Hilbert actions with additional matter contributions. It turns out that this procedure can be applied to brane action, which lead to a polynomial brane action.  We also examine the connection between this procedure and Polyakov trick in string theory.
		
		
%		PACS numbers: 04.50.Kd; 04.50.-h; 04.20.Fy
	\end{abstract}

\section{Introduction}
Nowadays the modification of gravity is seen by many as a promising way of solving various astrophysical and cosmological problems, e.g. the problem of dark matter and dark energy \cite{1106.2476,1504.04623}. It also serves as a starting point in gravity quantization \cite{1804.02262}.

One of the most popular method of gravity modification is the change of independent variables. New variables might include non-metric characteristics of geometry, additional fields etc. While such change itself does not necessarily lead to the modification of dynamics (the most famous example of this conservation of dynamics is the Hilbert-Palatini formalism), this modification can be achieved quite easily. For instance, if one makes the connection non-symmetric in the Hilbert-Palatini formalism, the new degrees of freedom appear immediately, which on a classical level can be associated with EM potential \cite{statja47}.

New variables might be connected with old ones (i.e. metric) in many ways. We will focus on the ways in which the metric can be expressed through derivatives of new variables. In this case dynamics changes due to the change of class of variations in variational principles (a toy model with this feature is discussed in \cite{statja33}).

The mimetic gravity is a new model which possesses this feature. It was proposed \cite{mukhanov} in an attempt to imitate the existence of dark matter: the modification of gravity is reduced to addition of dust matter with potential motion. Then it was immediately noticed \cite{Golovnev201439} that the action of the theory can be rewritten in the form of GR plus an additional {\emph{mimetic}} matter term which simplifies the analysis of its behavior. Later this action was generalized to the case of mimetic matter with arbitrary motion \cite{statja48}, for other generalizations see \cite{mimetic-review17}.

Another example of the theory with this feature which resembles a striking similarity with the mimetic gravity is the embedding theory, also known as Regge-Teitelboim (RT) approach \cite{regge}. In this approach the gravity is considered as a dynamics of 4-dimensional surface isometrically embedded in a flat ambient spacetime. The role of independent variable in this theory is played by embedding function which is connected with the metric by the condition that the metric on the surface is induced. This condition also involves
{differentiation}
of embedding function, so the extension of dynamics in this theory occurs due to the same reason that in the mimetic gravity. Moreover, the original RT action can be rewritten as an EH one plus some additional terms \cite{statja51}.

In the present paper we discuss a procedure that allows to rewrite any theory in which a metric is expressed through new variables, in the form of theory in which a metric is independent but the action contains an additional term. In the sections \ref{2} and \ref{3} we briefly review the
application of this procedure to the mimetic gravity and the embedding theory, where the additional action term is interpreted as some fictituous matter contribution. In the section \ref{4} we apply this procedure to a theory without the curvature, namely the theory of $n$-dimensional brane. In the section \ref{5} we show that this procedure is a natural generalization of the Polyakov trick.

\section{Mimetic Gravity}\label{2}
The mimetic gravity approach \cite{mukhanov} starts from the following redefinition of the metric:
\disn{v1}{
	g_{\m\n}=\tilde g_{\m\n}\tilde g^{\ga\de}(\dd_\ga\la)(\dd_\de\la),
	\nom}
where $\tilde{g}_{\mu\nu}$ is an auxiliary metric and $\la$ is an additional scalar field. This redefinition is often called the isolation of the conformal mode since the conformal transformation of the auxiliary metric does not affect the physical one, so auxiliary metric effectively has only nine independent components, and remaining one is parametrized by $\la$. These quantities form a set of independent variables in mimetic gravity. It is worth noting that the redefinition \eqref{v1} is a special case of disformal transformations, which were intensively studied in the context of the modification of gravity \cite{arXiv1407.0825}.

The action of mimetic gravity was originally taken as EH one in which the physical metric is expressed through new variables as in \eqref{v1}. The field equations of the theory are the following:
\disn{vi18}{
	G^{\m\n}=\ka \ls T^{\m\n}+n u^\m u^\n\rs,\qquad
	D_\m (n u^\m)=0,
	\nom}
where $T^{\m\n}$ is an EMT of ordinary matter, $D_\m$ is a covariant derivative, the signature is taken to be $(+---)$;
\begin{align}
n\equiv g_{\m\n}\ls \frac{1}{\ka}G^{\m\n}-T^{\m\n}\rs
\end{align}
is an effective density of the {\emph{mimetic matter}} particles,
\disn{vi17}{
	u_\m\equiv\dd_\m\la
	\nom}
is a four-velocity of it which obeys the usual normalization condition
\begin{align}\label{norm}
g^{\m\n}u_\m u_\n=1.
\end{align}
As can be seen, the field equations contains Einstein ones in which additional matter is present. One can thus raise a question: can this property be seen at the level of action, i.e. is it possible to rewrite the action of gravity as EH with independent metric plus additional {\emph{mimetic matter}} term?

Golovnev proposed \cite{Golovnev201439} the following form of the action:
\disn{v2}{
	S=-\frac{1}{2}\int\! d^4 x\, \sqrt{-g}\,\Bigl[R(g_{\m\n}) +n(1-g^{\m\n}(\dd_\m\la)(\dd_\n\la))\Bigr]=S_{EH}+S_{\text{add}},
	\nom}
in which the constraint \eqref{norm} is included with $n$ as a Lagrange multiplier. It can be noticed that $S_{\text{add}}$ is the action of pressureless ideal fluid with potential motion, so it can be rewritten as follows \cite{statja48}:
\disn{p2}{
	S_{\text{add}}=-\int\! d^4 x\, \sqrt{-g}\ls \sqrt{j^\m j^\n g_{\m\n}}-j^\m \dd_\m\la\rs,
	\nom}
where $j^\m$ is the mimetic current density and $\la$ is a Lagrange multiplier of the continuity equation of $j^\mu$. {The action $S_{\text{add}}$ can also be made}
polynomial w.r.t. $g_{\mu\nu}$ and mimetic matter variables by increasing their number \cite{statja48}:
\disn{p7}{
	S_{\text{add}}=-\frac{1}{2}\int\! d^4 x\, \ls  j^\m  j^\n g_{\m\n}+{a^2}-2a j^\m \dd_\m\la\rs,
	\nom}
where mimetic matter is described by  variables $(j^\m, \la, a)$. Note that here $j$ and $a$ differ from those in \cite{statja48} by the multiplier $(-g)^{1/4}$. One can construct many other forms of the action by integrating out some of these variables \cite{statja48}.

It is worth noting that the action \eqref{p7} can be generalized further to remove the potentiality restriction \eqref{vi17} without lost of polynomiality:
\disn{p71}{
	S_{\text{add}}=-\frac{1}{2}\int\! d^4 x\, \ls  j^\m  j^\n g_{\m\n}+{a^2}-2a j^\m (\dd_\m\la+\al \dd_\m \be)\rs,
	\nom}
where $\al$ and $\be$ are new scalar fields. However, in this case it becomes nonequivalent to original formulation, and the definition of physical metric which corresponds to the generalized action has to be altered \cite{statja48}:
\disn{v3}{
	g_{\m\n}=\tilde g_{\m\n}\tilde g^{\ga\de}(\dd_\ga\la+\al \dd_\ga \be)(\dd_\de\la+\al \dd_\de \be).
	\nom}

\section{Regge-Teitelboim Embedding Theory}\label{3}
In contrast with mimetic gravity, the embedding theory, originally proposed by Regge and Teitelboim \cite{regge}, has clear geometric sense. In this approach the gravity is described as a dynamics of surface isometrically embedded in a flat ambient space of $N$ dimensions. According to Friedman theorem \cite{fridman61}, in general case $N\geq 10$. The metric on the surface becomes induced and can be expressed through embedding function $y^a(x^{\mu})$ in the usual way:
\disn{r1}{
	g_{\m\n}=(\dd_\m y^a)(\dd_\n y^b)\,\eta_{ab},
	\nom}
where $a,b=0,\ldots,N-1$ and $\eta_{ab}$ is the ambient Minkowski metric.

The action of the theory is still EH action in which metric is assumed to be induced and $y^a$ is a set of independent variables. The equations of motion were originally written by Regge and Teitelboim in the following form:
\disn{r2}{
	D_\m\Bigl(( G^{\m\n}-\ka\, T^{\m\n}) \dd_\n y^a\Bigr)=0.
	\nom}
It is easy to see that they contain "extra" solutions for which $G^{\m\n}\neq\ka\, T^{\m\n}$. This property was initially treated as a drawback of the theory \cite{deser}, but eventually it become a starting point in search for the explanation of dark matter \cite{davkar,statja26}.

The presence of an additional matter can be noticed when one tries to rewrite the RT equations in the form similar to \eqref{vi18}, i.e. in the form of Einstein equations with additional matter\cite{pavsic85,statja33}:
\disn{r3}{
	G^{\m\n}=\ka \ls T^{\m\n}+\ta^{\m\n}\rs,\qquad
	\nom}
where $\tau^{\mu\nu}$ can be interpreted as EMT of additional matter which dynamics is governed by equation
\disn{r4}{
	D_\m\Bigl(\ta^{\m\n}\dd_\n y^a\Bigr)=0.
	\nom}

We then raise the same question as in the previous section: how to write the contribution of gravity to the full action of the theory in the form
\begin{align}
S=S_{EH}+S_{\text{add}},
\end{align}
i.e. as a EH action with an independent metric plus an additional {\emph{embedding matter}} term?
The most straightforward way to do it is to choose $S_{add}$ as a constraint \eqref{r1} with a Lagrange multiplier $\ta^{\m\n}$ \cite{statja48}:
\disn{za1}{
	S_{\text{add}}=\frac{1}{2}\int\! d^4 x\, \sqrt{-g}\,\Bigl( (\dd_\m y^a)(\dd_\n y_a) - g_{\m\n}\Bigr)\tau^{\m\n}.
	\nom}
Note that in this case, as in mimetic gravity, the gravitational action consists of EH one and an additional contribution which is a constraint arising from the substitution for the physical metric (it is easy to check that the constraint \eqref{norm} is a corollary of the substitution \eqref{v1} and the notation \eqref{vi17}).

The other actions  mentioned in the previous section also have analogs in the embedding theory. Namely, the action of embedding theory which corresponds to the action \eqref{p2} can be obtained through increasing the number of currents \cite{statja51}:
\disn{r5}{
	S_{\text{add}}=\int\! d^4 x\, \sqrt{-g}\,
	\Bigl( j^\m_a\dd_\m y^a-\text{\bf tr}\sqrt{g_{\m\n}j^\n_a j^{\al a}}\Bigr),
	\nom}
where $j^{a\mu}$ is a set of currents corresponding to embedding matter and $\sqrt{\;\;}$ denotes the square root of matrix. {The action $S_{\text{add}}$ can also be made}
polynomial w.r.t. $g_{\mu\nu}$ and embedding matter variables  as it was done in  \eqref{p7} by increasing the number of independent variables \cite{statja51}:
\disn{r14}{
	S_{\text{add}}=\int\! d^4 x\,
	\biggl( j^\m_a\dd_\m y^a-\be^{\m\n}g_{\m\n}+\frac{1}{2}\la_{\al\be}\ls\be^{\al\ga}g_{\ga\de}\be^{\de\be}-j^\al_a j^{\be a}\rs\biggr),
	\nom}
where $\be^{\m\n}$ and $\la_{\al\be}$ are symmetric tensor fields. Note that $j$, $\be$ and $\la$ differ from used in \cite{statja51} by a multiplier like $\sqrt{-g}$.

Note that
due to \eqref{vi17} the motion of mimetic matter in the original formulation is governed by only one scalar function $\lambda$, so one has to extend this formulation to allow more general motion. To the contrary, the motion of the embedding matter is governed by a set of fields which does not restrict its motion as much as $\la$ in mimetic gravity. Therefore the main problem here is not the extension of the dynamics, but rather the search for the anzats which leads to the cases of matter with physically interesting properties.

\section{Brane}\label{4}
In the previous sections we applied the procedure, which allows to replace the substitution of metric through new variables with the addition of a certain term to the action of modified gravity theories. However, it turns out that such procedure can be of use in other theories as well. For example, it is not necessary to have the curvature term in the action to apply this procedure, which, as we have seen above, essentially has the following steps:
\begin{itemize}
	\item We start with an action which contains metric expressed through derivatives of some other variables.
	\item Then we make metric independent again and add the constraints, which relate it to the other variables, to the action with Lagrange multipliers.
	\item The resulting additional terms can often be simplified by changing the number and the type of independent variables.
\end{itemize}

Let us consider a simpler action, namely the action of $n$-dimensional brane:
\disn{za0}{
	S=-T\int\! d^n x\, \sqrt{-g^{\text{ind}}},
	\nom}
where $g^{\text{ind}}$ is the determinant of the induced metric \eqref{r1}.
The application of the above procedure gives
\disn{za3}{
	S=-T\int\! d^n x\, \,
	\Bigl(\sqrt{-g}+ \frac{\tau^{\m\n}}{2}\bigl((\dd_\m y^a)(\dd_\n y_a) - g_{\m\n}\bigr)\Bigr).
	\nom}
It is easy to notice that this action can be made polynomial by the introduction of
standard ($g_{\m\n}=e^A_{\mu}e^B_{\nu}\eta_{AB}$) vielbein $e^A_{\mu}$:
\disn{za2}{
	S=-T\int\! d^n x\,\,
	\Bigl( e+ \frac{\tau^{\m\n}}{2} \bigl((\dd_\m y^a)(\dd_\n y_a) - e_\mu^A e_{A\nu}\bigr)\Bigr),
	\nom}
where $A=0,\ldots,n-1$; $e=\det e^A_{\mu}$. While polynomial action for the bosonic brane was discussed in the literature \cite{guendelman,0909.4151},  it is worth noting that the action \eqref{za2} can be obtained in a quite simple way. Moreover, the action \eqref{za3} can be simplified further in the case when $n=2$ which is the subject of the next section.

\section{String}\label{5}
Let us consider the classical Nambu-Goto action for a bosonic string:
\disn{za3a}{
	S=-T\int\! d^2 x\, \sqrt{-g^{\text{ind}}}.
	\nom}
Then we can immediately write down the 2-dimensional analog of \eqref{za3}:
\disn{za3b}{
	S=-T\int\! d^2 x\, \,
	\Bigl(\sqrt{-g}+ \frac{\tau^{\m\n}}{2}\bigl((\dd_\m y^a)(\dd_\n y_a) - g_{\m\n}\bigr)\Bigr).
	\nom}
The remarkable feature of $2D$
case is that the further simplification
can be achieved not by raising but rather by lowering the number of independent variables. Namely, $g_{\mu\nu}$ can be integrated out (we did not consider that possibility in the sections \ref{2} and \ref{3} because we wanted to keep EH term in the resulting theory).

Indeed, the variation w.r.t. $g_{\mu\nu}$ gives
\begin{align}\label{tau}
\sqrt{-g}g^{\m\n}=\tau^{\m\n}.
\end{align}
Taking the determinant of this equation and noting that $\det(\sqrt{-g}g^{\m\n})=-1$,
we obtain the restriction for $\tau^{\m\n}$:
\disn{r24}{
	\det \tau^{\m\n}=-1.
	\nom}
On the other side, since the left-hand side of the equation \eqref{tau} is invariant w.r.t. scaling of metric,
we can find $g_{\m\n}$ from \eqref{tau} only up to a multiplier:
\disn{r23}{
	g_{\m\n}=\alpha\tau^{-1}_{\m\n},
	\nom}
where $\al$ is an arbitrary function.
However, it turns out that we are still able to exclude $g_{\m\n}$ from the action completely, since after the substitution of \eqref{r23} into \eqref{za3b} assuming \eqref{r24} the
first and third terms cancel each other (this takes place only at $n=2$), and the result does not depend on $\al$:
\disn{za3b2}{
	S=-\frac{T}{2}\int\! d^2 x\, \tau^{\m\n}(\dd_\m y^a)(\dd_\n y_a).
	\nom}

To satisfy the restriction \eqref{r24} one can introduce a new auxiliary metric $h_{\m\n}$ and rewrite $\tau^{\mu\nu}$ through it:
\begin{align}\label{hmn}
\tau^{\m\n} = \sqrt{-h}\,h^{\m\n}.
\end{align}
As a result, the action \eqref{za3b2} takes the form
\disn{r25}{
	S=-\frac{T}{2}\int\! d^2x\,\sqrt{-h}\,h^{\m\n} (\dd_\m y^a)(\dd_\n y_a),
	\nom}
which is the famous Polyakov action.

\section{Discussion}
The main topic of interest in the present paper is the transformation of action for various geometric theories.

Besides the case of mimetic gravity, which is quite widely discussed in the literature, we consider the the action of the world sheet in $D$ dimensions. We also consider the similar case of the EH action in which the metric is assumed to be induced \eqref{r1} (gravitation {\emph{a la string}}). 
In a result of transformation a new expression for the action appears, which is a sum of the original action with independent metric and some additional term.

However, an analogous approach can be used for other modifications of GR in which the extension of dynamics occurs.  The extension appearing in $f(R)$ is especially interesting. Note that, in contrast with theories considered in the present paper, one obtains this theory not by the substitution of expression for metric through new variables in the action, but by the change of the form of the action itself. Nevertheless, it is well known that it can also be obtained when one considers the sum of the original EH action and the additional contribution corresponding to the scalar field with some potential (see, e.g. \cite{gr-qc/0604028}). In a result of this, as well as in theories considered in the present paper, certain gravitational degrees of freedom transform into degrees of freedom of some additional matter.
It might simplify the analysis of solutions and can be of use in the cosmological applications,
e.g., in the study of the cosmological singularities \cite{0903.2753},
or in search for explanations of  accelerated expansion of the Universe \cite{gr-qc/0201033}. 
Of particular interest are cases in which the additional matter turn out to be perfect fluids \cite{1810.03204}. Furthermore, the generalizations on the case of $f(R,\cal{G})$ gravity, where $\cal{G}$ is the Gauss-Bonnet topological invariant, are also possible \cite{1906.05693}.

\section*{Acknowledgments}
The work of A.~S. was supported by RFBR Grant No.~18-31-00169.

\end{document}